\documentclass[sigconf]{acmart}
\AtBeginDocument{%
  }

\usepackage{tabularx}
\usepackage{booktabs}
\usepackage{multirow}
\usepackage{enumitem}
\setcopyright{acmlicensed}
\copyrightyear{2025}
\acmYear{2025}
\acmDOI{XXXXXXX.XXXXXXX}

\acmConference[Conference acronym 'XX]{}{June 03--05,
  2018}{Woodstock, NY}
\acmISBN{978-1-4503-XXXX-X/2018/06}

\begin{document}

\title{SessionRec: Next Session Prediction Paradigm For Generative Sequential Recommendation}

\author{Lei Huang}
\email{huanglei45@meituan.com}
\author{Hao Guo}
\email{guohao15@meituan.com}
\affiliation{%
  \institution{Meituan}
  \city{Beijing}
  \country{China}
}

\author{Linzhi Peng}
\authornote{This work was done during the internship at Meituan.}
\email{lzpeng626@buaa.edu.cn}
\affiliation{%
  \institution{Beihang University}
  \city{Beijing}
  \country{China}}

\author{Long Zhang}
\email{zhanglong40@meituan.com}
\author{Xiaoteng Wang}
\email{wangxiaoteng03@meituan.com}
\affiliation{%
  \institution{Meituan}
  \city{Beijing}
  \country{China}
}

\author{Daoyuan Wang}
\email{wangdaoyuan@meituan.com}
\author{Shichao Wang}
\email{wangshichao10@meituan.com}
\affiliation{%
  \institution{Meituan}
  \city{Beijing}
  \country{China}
}

\author{Jinpeng Wang}
\email{wangjinpeng04@meituan.com}
\author{Lei Wang}
\email{wanglei46@meituan.com}
\affiliation{%
  \institution{Meituan}
  \city{Beijing}
  \country{China}
}

\author{Sheng Chen}
\email{chensheng19@meituan.com}
\affiliation{%
  \institution{Meituan}
  \city{Beijing}
  \country{China}
}

\renewcommand{\shortauthors}{Lei et al.}

\begin{abstract}

We introduce SessionRec, a novel next-session prediction paradigm (NSPP) for generative sequential recommendation, addressing the fundamental misalignment between conventional next-item prediction paradigm (NIPP) and real-world recommendation scenarios. Unlike NIPP's item-level autoregressive generation that contradicts actual session-based user interactions, our framework introduces a session-aware representation learning through hierarchical sequence aggregation (intra/inter-session), reducing attention computation complexity while enabling implicit modeling of massive negative interactions, and a session-based prediction objective that better captures users' diverse interests through multi-item recommendation in next sessions. Moreover, we found that incorporating a rank loss for items within the session under the next session prediction paradigm can significantly improve the ranking effectiveness of generative sequence recommendation models. We also verified that SessionRec exhibits clear power-law scaling laws similar to those observed in LLMs. Extensive experiments conducted on public datasets and online A/B test in Meituan App demonstrate the effectiveness of SessionRec. The proposed paradigm establishes new foundations for developing industrial-scale generative recommendation systems through its model-agnostic architecture and computational efficiency.

\end{abstract}




\maketitle

\begin{figure}[h]
\vspace{-0.5cm}
  \centering
  \includegraphics[width=\linewidth]{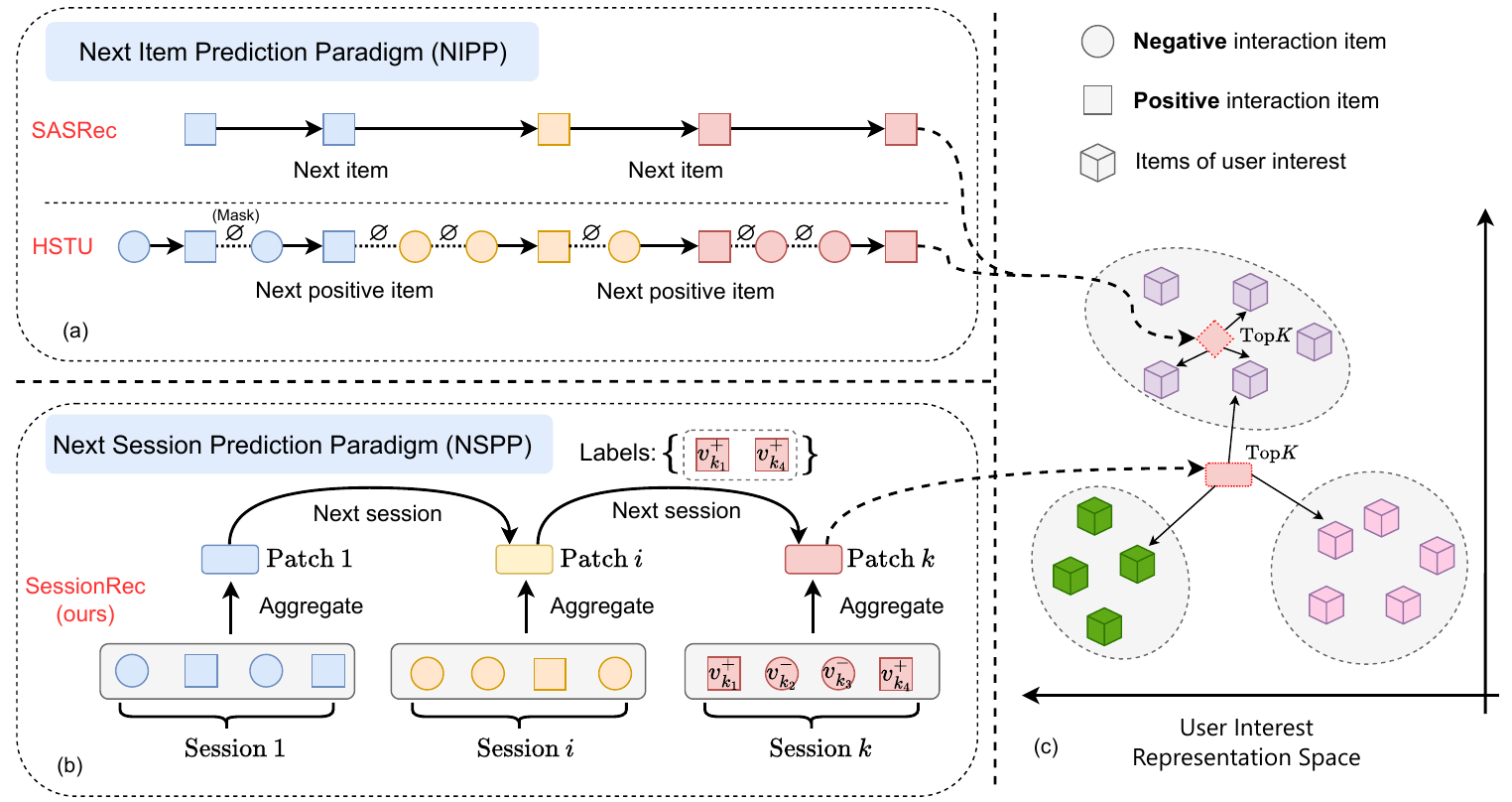}
  \vspace{-0.5cm}
  \caption{Next Item Prediction Paradigm (NIPP) vs. our proposed Next Session Prediction Paradigm (NSPP).}
 \vspace{-0.5cm}
  \label{fig:overview}
\end{figure}

\section{Introduction}
Large Language Models (LLMs) \cite{radford2019gpt2, brown2020gpt3, achiam2023gpt4, ouyang2022training, wei2021finetuned, kaplan2020scaling} have exhibited tremendous success across various fields due to their highly scalable and generalizable nature. The core strategy of LLMs, particularly the GPT series \cite{radford2018gpt1, radford2019gpt2, brown2020gpt3, achiam2023gpt4}, involves predicting the next token by encoding a sequence of input tokens, a process integral to both model training and text generation.
Notably, similar paradigms exist within generative sequential recommendation models. For instance, GRU4Rec \cite{hidasi2018gru4rec}, SASRec \cite{kang2018self}, and BERT4Rec \cite{sun2019bert4rec} leverage comparable strategies by predicting subsequent item in users' interaction histories. 
These models mirror the scalability and generalizability advantages of LLMs, thereby enhancing the performance of recommendation systems.
Meta's recent introduction of HSTU, a unified generative recommendation framework \cite{zhai24hstu}, demonstrates substantial advantages in industrial scenarios and corroborates the scaling laws within generative sequential recommendations.

Current generative sequential recommendation models, following the Next Item Prediction Paradigm (NIPP), treat tokens as items and recommend user interests through an autoregressive, item-by-item process. 
However, this approach diverges from real-world needs, where user interactions occur at the session/request level, with all items generated simultaneously when the session initiates. Thus, sessions, rather than individual items, should form the basic unit in recommendation systems.
Furthermore, there is a misalignment with the NIPP's modeling objectives as existing models often predict only one item per user session. In contrast, user interests are diverse, necessitating the prediction of multiple items that the user might be interested in during their next session.
Additionally, models like HSTU leverage full historical item interactions (e.g., exposure, click, like, dislike) to refine user interest representations and other models also emphasize the importance of cross-statistical features \cite{zhai24hstu,cheng2016wide, lian2018xdeepfm, song2019autoint}. Although incorporating users' negative interactions for the extraction of implicit statistical features has been shown to enhance recommendations, the reliance on training primarily with positive interactions or using sampled cross-entropy loss for the next positive item \cite{sun2019bert4rec, klenitskiy2023turning, tang2018personalized} constrains effective item ranking and accurate predictions within sessions.

To address these considerations, we introduce the \textbf{Next Session Prediction Paradigm} (NSPP) in generative sequential recommendation models. As illustrated in Figure \ref{fig:overview}, NSPP differs markedly from the traditional NIPP.
Firstly, \textbf{the user representation paradigm} adopts session-level embeddings with both positive and negative interactions to capture broader contextual and cross statistical information within user sessions, which is inspired by visual transformers in computer vision \cite{liu2021swin, dosovitskiy2020image} that segment images into patches instead of single pixels.
Secondly, \textbf{the joint learning paradigm} aligns both retrieval and ranking tasks concurrently within a session, utilizing comprehensive interaction data.
Finally, \textbf{the label prediction paradigm} predicts positive interactions in the next session to accommodate several potentially intriguing items for the user. 

Overall, we propose a general NSPP framework for generative sequential recommendation, which is model-agnostic with plug-and-play characteristics. It can be tailored to integrate with any sequence encoding backbone such as RNN \cite{chung2014gru, yu2019review}, the conventional Transformer \cite{waswani2017transformer} and HSTU \cite{zhai24hstu}.
It also enhances sequence encoding efficiency and reduces computational complexity. In Transformer-like architectures \cite{waswani2017transformer}, computational complexity scales as
\(O(N^2)\) with input sequence length. By employing intra-session and inter-session sequence aggregation, NSPP significantly lowers computational demands, especially when incorporating a large amount of negative interaction behavior into the input sequence. It benefits both offline training and online serving in industrial recommendation systems.
With the session of comprehensive interactions, we develop a rank loss to distinguish the negative behaviors in the prediction session, enabling more fine-grained predictions.
This strengthens our conviction that a single generative model can be employed to manage both retrieval and ranking tasks concurrently in the next session paradigm, effectively replacing the traditional recommendation system's multi-model cascaded approach of retrieval, pre-ranking, and ranking. 
Last but not least, we also verified that the next session prediction paradigm continues to demonstrate scaling-law characteristics, with performance continuously improving as the data volume increases.

Our contributions are summarized as follows:
\vspace{-0.1cm}
\begin{itemize}[leftmargin=\parindent]
    \item  We propose \textbf{SessionRec}, a novel next session prediction paradigm for generative sequential recommendation, by reconsidering how users engage with recommendation systems in practical applications, offering new insights into the design of generative recommendation algorithms.
    \item We further propose a rank loss that captures the model's intra-session ranking capabilities, which can significantly enhance the ranking performance of generative recommendation models.
    \item Extensive experiments conducted on public datasets and A/B test in online recommendation system of Meituan App demonstrate the effectiveness of our proposed SessionRec, notably with average 27\% performance gains over the best baseline on the public datasets.

\end{itemize}

\section{Related Work}
\textbf{Behavior Sequence Modeling.} 
In recommendation systems, user behavior sequences contain a wealth of valuable information that reflects user interest preferences, which is crucial to characterizing user personalization. Several studies leverage users' historical interaction behavior sequences to understand their preferences and recommend items that they might be interested in.
DIN \cite{zhou2018din}, DIEN \cite{zhou2019dien}, DSIN \cite{feng2019dsin}, and MIMN \cite{pi2019mimn} introduced the attention mechanism in modeling behavior sequences. By calculating the correlation between user historical behaviors and the target item, it assigns a dynamic weight to each behavior to better capture user interests. 
While longer user behavior sequences can provide more useful information for modeling user interests, they also place a significant burden on the latency and storage requirements of online serving systems. An possible solution is to retrieve the most relevant and important behaviors from an extremely long sequence by matching algorithm, such as category id \cite{pi2020search}, locality-sensitive hashing (LSH) \cite{chen2021end}, SimHash\cite{cao2022sampling}, and Efficient-Target-Attention \cite{chang2023twin}. \\

\noindent \textbf{Sequential Recommendation.}
Sequential recommendation models treat recommendation as a sequence-to-sequence generation task according to Markov Chains assumption.
They generate the next item directly by considering the sequence of past user interactions. 
Methods such as GRU4Rec \cite{hidasi2018gru4rec} and LSTM-based approaches \cite{wu2017recurrent} were particularly effective in capturing both long-term and short-term patterns in item transitions. Following the success of self-attention mechanisms \cite{waswani2017transformer} in natural language processing (NLP), researchers developed many Transformer-based models for sequential recommendation. SASRec \cite{kang2018self} applies self-attention to sequential recommendation tasks, while BERT4Rec\cite{sun2019bert4rec} uses the BERT architecture to model bidirectional relationships between items in a sequence. However, given the sparsity of user behaviors and the computational complexity, these methods typically focus only on users' positive behaviors in industrual implementations.
HSTU \cite{zhai24hstu} incorporates both positive and negative user behaviors into the sequence, capturing more accurate interests and preferences. 
\\

\noindent \textbf{Large Language Models for Recommender Systems.} Following the advent of ChatGPT, there has been an uptick in efforts within both the industry and academia to integrate Large Language Models (LLMs) into recommendation systems. Some studies utilize LLMs to produce semantic embeddings for items, which are then converted into semantic IDs through algorithms such as RQ-VAE \cite{lee2022autoregressive, rajput2023recommender}. UEM \cite{doddapaneni2024user} processed user history as plain text, generating token embeddings for history items. This approach greatly simplified user history tracking and enabled the incorporation of longer user histories into the language model, and allowed their representations to be learned in context.
ILM \cite{yang2024item} incorporating collaborative filtering knowledge into a frozen LLM for conversational recommendation tasks.
LLMs can also be used to encode target feature entities \cite{chen2024hllm}, extracting their implicit embedding representations to feed into subsequent recommendation models. 
These methods primarily leverage the content understanding capabilities of LLMs to provide additional informational gains, rather than applying the underlying paradigm of LLM modeling directly to the recommendation system.

\section{Next Session Prediction Paradigm}
\subsection{Problem Statement}
In sequential recommendation, let \begin{math}\mathcal{U} = \{u_1, u_2, \ldots, u_{|\mathcal{U}|}\} \end{math} denote a set of users, \begin{math} \mathcal{V} = \{v_1, v_2, \ldots, v_{|\mathcal{V}|}\} \end{math} be a set of items, and each item \begin{math} v \end{math} is associated with some side information (e.g., action type, brand) \begin{math} F_v = \{f_1, f_2, \ldots, f_c\} \end{math}, where \begin{math} c \end{math} is the number of side information features. For each user, the historical interaction sequence can be represented as \begin{math}
\mathcal{S}_u = [(v_1, F_1, s_1), (v_2, F_2, s_2), \ldots, (v_n, F_n, s_n)]\end{math}, where \begin{math} n \end{math} denotes the sequence length, and the value of \begin{math} s_i \end{math} denotes the session identifier. If the values of \( s_i \) are equal, it indicates that the interactions belong to the same session. Note that the historical interactions include not only positive interactions (e.g., click, purchase, like), but also negative interactions (e.g., exposure, dislike). 

Given the interaction sequence, NSPP aims to predict the items that user \begin{math} u \end{math} will interact positively with in the next user session, which can be formulated as: 
\begin{equation}
    [v_{n+1}, v_{n+2}, \ldots, v_{n+k} ] = \textbf{NSPP}({S}_u; SeqEncoder),
\end{equation}
where \( k \) denotes the number of positive interaction items in next session. \begin{math} SeqEncoder \end{math} denotes the sequence encoding backbone with learnable parameters. While the NIPP aims to predict the next positive interaction item, which can be formulated as:
\begin{equation}
    v_{n+1} = \textbf{NIPP}({S}_u; SeqEncoder).
\end{equation}
It is evident that NSPP aligns more closely with the objectives of online recommendation systems. As a user session/request arrives, all items that the user might be positive interacted with need to be predicted at once, and it is impossible to predict item by item.

\begin{figure*}[t]
  \centering
  \includegraphics[width=0.95\textwidth]{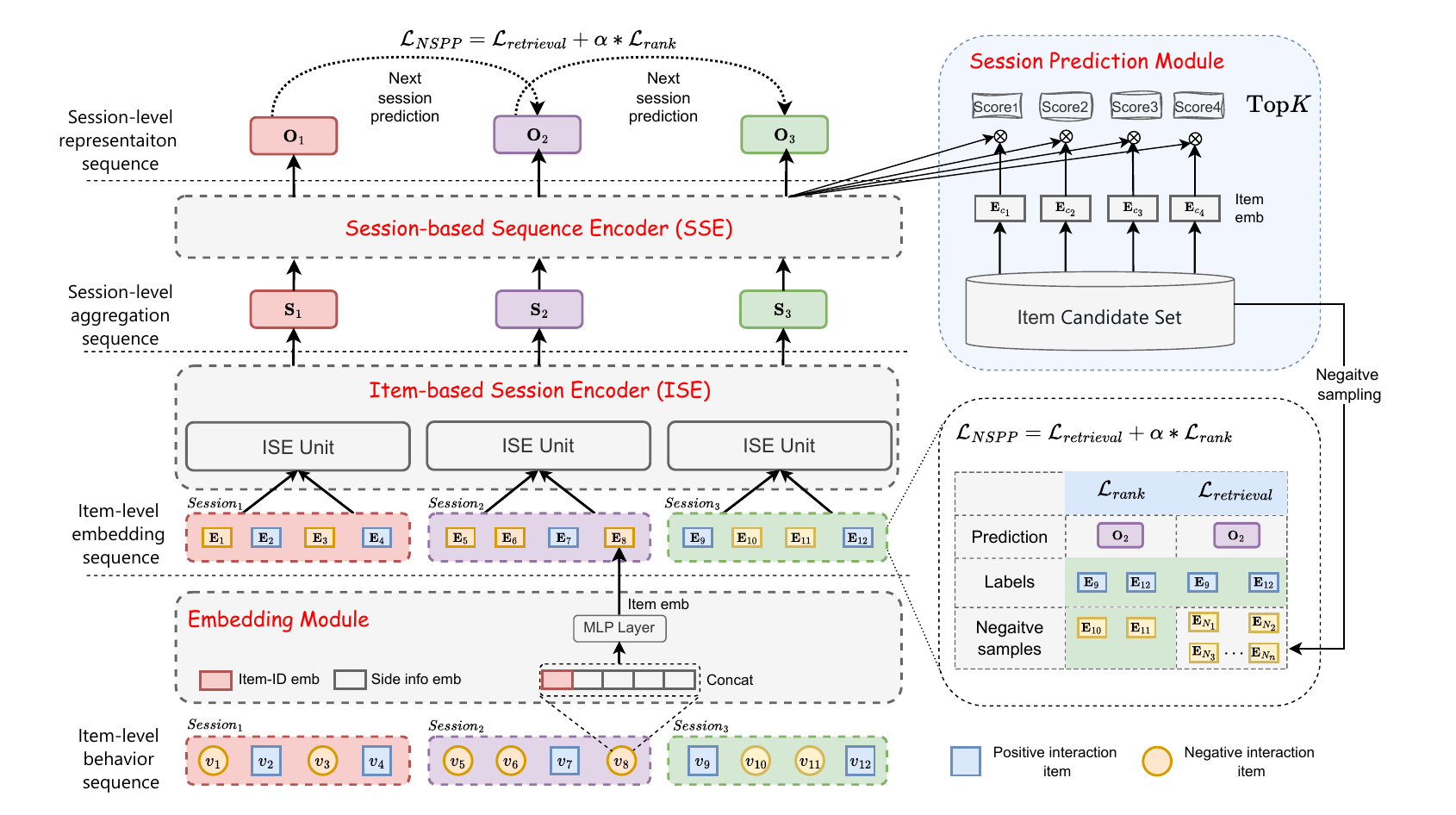}
    \vspace{-0.3cm}
  \caption{Overall architecture of our proposed SessionRec.}
  \label{fig:framework}
  \vspace{-0.5cm}
\end{figure*}

\subsection{The Overall Framework}

The overall framework of SessionRec is shown in Figure \ref{fig:framework}.  SessionRec is a hierarchical encoding structure that performs intra-session encoding followed by inter-session encoding, which is composed of four components: Embedding Module, Item-based Session Encoder (ISE), Session-based Sequence Encoder (SSE), and Session Prediction Module. ISE aggregates and encodes intra-session information to obtain session representations, which serve as the tokens of the session sequence. SSE then encodes the session sequence to derive the user interest representation.
During this forward computation, the user's behavior sequence transforms from an item-level sequence to a session-level sequence. 
In the session prediction module, both the user representation and the prediction label are session-level.
This approach also helps reduce the computational complexity of the network and effectively models the user's multiple interests.

\subsection{Embedding Module}
The model takes the historical user interaction sequence as input, where the item-ID and the item side information are discrete features. When side information comprises continuous attributes, techniques such as binning \cite{liu1998integrating, zhang1996birch} can be used to discretize them. For each discrete feature, an embedding table \begin{math} \mathbf{X}_f \in \mathbb{F}^{|\mathcal{X}| \times d_f} \end{math} can be assigned to store embeddings, where \( d_f \) is the embedding dimension. The feature embedding of value \( f_i\) can be obtained through a look-up table operation:
\begin{equation}
    \textbf{E}_{f_i} = \text{LookUpTable}(\mathbf{X}_f, f_i).
\end{equation}
The item-ID embeddings \(\textbf{E}_{iid}\) and the side information embeddings are then concatenated and transformed into the item embedding of dimension \(d\) through the MLP layer, which is formulated as:
\begin{equation}
    \textbf{E}_v = \text{MLP}([\textbf{E}_{iid} || \textbf{E}_{f_1}|| \textbf{E}_{f_2}|| \ldots || \textbf{E}_{f_c}]).
\end{equation}
where \(||\) denotes the vector concatenation operation.

\subsection{Item-based Session Encoder}
The interaction item sequence is transformed to the item embedding sequence as \begin{math}  \mathcal{S}_u^{'} =  [(\textbf{E}_{v_1}, s_i), (\textbf{E}_{v_2},s_2),  \ldots , (\textbf{E}_{v_n}, s_n)] \end{math} after embedding module. The Item-based Session Encoder (ISE) encodes and aggregates all item embedding sequences corresponding to each session identifier, which transforms the item-level embedding sequence into a session-level representation sequence as:
\begin{equation}
    [\textbf{ES}_1, \textbf{ES}_2, \ldots, \textbf{ES}_m] = \text{ISE}([(\textbf{E}_{v_1}, s_1), (\textbf{E}_{v_2},s_2),  \ldots , (\textbf{E}_{v_n}, s_n)] ),
\end{equation}
where \(\textbf{ES}\) denotes the session representation output by ISE, and \(m\) denotes the number of sessions. In real-world scenarios, the number of items within each session may vary. To manage this during implementation, one can either use padding strategy to standardize session lengths or utilize RaggedTensor\footnote{\url{https://www.tensorflow.org/api_docs/python/tf/RaggedTensor}} for flexibility.

ISE can employ parameter-free pooling methods (such as max pooling or mean pooling) or parameterized methods like RNNs, Transformers. Our best practice is to employ the pooling method for ISE because it is straightforward and effective, achieving performance comparable to those of more complex encoders. In the experimental section, we performed an in-depth analysis of how various session encoders influence performance.

\subsection{Session-based Sequence Encoder}

The Session-based Sequence Encoder (SSE) takes the session-level representation sequence obtained from the Item-based Session Encoder (ISE) as input and encodes the entire sequence to derive the final user interest representation, which is the last output of the sequence encoder. This can be formulated as:
\begin{equation}
    \textbf{E}_u = \text{SSE}([\textbf{ES}_1, \textbf{ES}_2, \ldots, \textbf{ES}_m] ),
\end{equation}
where \(\textbf{E}_u\) represents the encoded user interest representation. 

Since our proposed NSPP is model-agnostic, SSE can be tailored to integrate with any sequence encoding backbone, including, but not limited to, RNN \cite{chung2014gru, yu2019review}, the conventional Transformer \cite{waswani2017transformer} and HSTU \cite{zhai24hstu}. For improved modeling of interaction information among session-level tokens and to prevent temporal leakage issues during sequence encoding, a decoder-only Transformer architecture is advised for the SSE. This could be a traditional Transformer or an HSTU. Our best practice for SSE is to implement HSTU to achieve the best performance.

In our proposed NSPP, the input to the sequence encoder is transformed from an item-level sequence to a session-level sequence. Assuming the average session length is \( M \) (with \( M=23 \) in the Meituan homepage recommendation scenario), the length of the input sequence to the sequence encoder is reduced by an average factor of \( M \). If a Transformer-like architecture is used as the backbone, the computational complexity of the sequence encoder will be reduced by a factor of \( M^2 \). This reduction offers significant benefits for both model training and online inference.

\subsection{Session Prediction Module}

The user interest representation embedding derived from the SSE encodes information about items that the user is likely to engage with positively in the upcoming session. Therefore, the likelihood of a user having a positive interaction with an item \( v_i \), denoted as \( P_u^{v_i} \), is indicated by the similarity score between the user representation embedding \( \textbf{E}_u \) and the item embedding \( \textbf{E}_{v_i} \):
\begin{equation}
    P_u^{v_i} = \text{DotProduct}(\textbf{E}_u, \textbf{E}_{v_i}),
\end{equation}
where \(\text{DotProduct}\) represents the commonly useded operation of calculating the similarity between vectors through their dot product. The top \( K \) items with the highest similarity scores constitute the recommended results output by the model. In real-world industrial applications, given the extensive quantity of items that exceed hundreds of millions, the Approximate Nearest Neighbor (ANN) technique \cite{malkov2018ann} is utilized to efficiently identify the top \( K \) items most related to the user interest representation.

\subsection{Network Training}

\subsubsection{Retrieval loss.}  
In training a model based on NSPP, the prediction target for each position's output in the sequence encoder involves the information of all positively interacted item tokens in the next session. We adopt the widely used sampled cross-entropy (CE) loss \cite{hidasi2018gru4rec, klenitskiy2023turning} to train our model:
\begin{equation}
    \mathcal{L}_{retrieval} = - \sum_{i=1}^{N-1} \sum_{v_j \in \mathcal{V}_{s_i}^+} \text{log} \frac{\exp(P_{s_i}^{v_j})}{\exp(P_{s_i}^{v_j}) + \sum_{v_k \in \mathcal{V}_C^-} \exp(P_{s_i}^{v_k})},
\label{eq:loss_recall}
\end{equation}
where \( N \) denotes the length of the user's session-level sequence,  \( \mathcal{V}_{s_i}^+\) denotes the set of positive interaction items in the next session to be predicted, and \( P_{s_i}^{v_j} \) represents the dot product similarity between the output vector at the \( i \)-th position of the sequence decoder and the item embedding of item \( v_j \). \( \mathcal{V}_C^-\)  denotes a random negative sampling set of size \( C \). This is because, in practical scenarios, the item set can be exceedingly large, reaching hundreds of millions. Therefore, random negative sampling is used during training to avoid heavy computation.

The optimization objective of this loss function is to select the target item from the set of candidate items, which aligns with the retrieval task. Thus, we name it the retrieval loss.

\subsubsection{Rank loss.}

To further enhance the model's ability to distinguish hard negative samples, we introduce a more challenging optimization objective on top of the retrieval loss. This means that the negative samples are not randomly sampled items but are instead obtained from negatively interacted items within the next session. The optimization objective is to rank items with which the user has positive interactions ahead of those with negative interactions within the same session. This aligns with the goal of the ranking task in recommendation systems, and thus we refer to it as the rank loss: 
\begin{equation}
    \mathcal{L}_{rank} = - \sum_{i=1}^{N-1} \sum_{v_j \in \mathcal{V}_{s_i}^+} \text{log} \frac{\exp(P_{s_i}^{v_j})}{\exp(P_{s_i}^{v_j}) + \sum_{v_k \in \mathcal{V}_{s_i}^-} \exp(P_{s_i}^{v_k})},
\label{eq:loss_rank}
\end{equation}
where \( \mathcal{V}_{s_i}^-\) represents the set of negative interaction items within the next session to be predicted. 

\subsubsection{Balance retrieval loss and rank loss.}

We multiply the rank loss by a weighting coefficient \(\alpha\) to balance the retrieval loss and the rank loss, enabling the model to possess both retrieval and ranking capabilities. The final loss function of NSPP is formulated as:
\begin{equation}
    \mathcal{L}_{NSPP} = \mathcal{L}_{retrival} + \alpha * \mathcal{L}_{rank}.
\label{eq:loss_fianl}
\end{equation}

Analogously to the cascaded paradigm in recommendation systems where retrieval is followed by ranking, it is meaningful to first ensure that positive samples can be retrieved before optimizing ranking capabilities. Our experiments have validated this perspective. Specifically, incorporating a relatively small rank loss in the loss function can lead to substantial improvements in ranking performance. However, if the weight of the rank loss is set too high, the model may neglect retrieval capabilities during training, which can ultimately damage overall model performance.

\section{Experiments}

\subsection{Experimental Setting}
\subsubsection{Datasets}

Although datasets such as\textit{ MovieLens (1M, 20M)} \cite{harper2015movielens} and \textit{Amazon series} \cite{mcauley2015amazon} have been widely used in previous work, they lack session identifier information, which is essential for session aggregation. Additionally, these datasets only provide items that users have positively interacted with. Nevertheless, negative interactions, or exposure behaviors, are very crucial for implicit feature extraction in industry scenarios. Therefore, we employ two public datasets from real-world applications that better correspond to the data from actual recommendation systems. The statistics of the datasets are shown in Table 1. 
\begin{table}[t]
\caption{Statistics of datasets.}
\vspace{-0.3cm}
\label{tab:Notations}
\centering
\renewcommand{\arraystretch}{1.2} % 调整行间距
\setlength{\tabcolsep}{4pt} % 调整列间距
\footnotesize
\begin{tabular}{c c c c c p{1.5cm}}
\toprule
\textbf{Datasets} & \textbf{\#users} & \textbf{\#items} & \textbf{\#interactions} & \textbf{Avg.length} & \textbf{Avg.positive length} \\
\midrule
KuaiSAR & 9,460 & 109,575 & 758,560 & 80.19 & 15.44 \\
RecFlow & 28,257 & 1,482,117 & 22,896,181 & 810.28 & 404.11 \\
% Meituan & 100,000 & 5,423,549 & 165,018,494 & 1,650.18 & 117.02 \\
\bottomrule
\end{tabular}
% \vspace{-0.5cm}
\end{table}

\textbf{KuaiSAR-S\footnote{\url{https://kuaisar.github.io/}}:} This dataset contains genuine user behavior logs from Kuaishou, a leading short video application in China with over 400 million daily active users. It captures detailed search interactions of 25,877 users over 19 days, providing a comprehensive view of user engagement.

\textbf{RecFlow\footnote{\url{https://github.com/RecFlow-ICLR/RecFlow}}:} 
This is an industrial dataset containing positive and negative samples from the exposure space, along with effective view data for positive feedback. It is particularly useful for user behavior sequence analysis.

For dataset preprocessing, we treat interactions with only exposure behaviors as negative interactions, while those with effective views, click or purchase behaviors are considered positive interactions. Following common practices, we retain items and users with at least five feedbacks, sessions with at least one positive sample, and users with a minimum of three sessions.
\begin{table*}[t]
\centering
\caption{Overall Performance Comparison. The best and runner-up results are bold and underlined, respectively. “Impv.” means the improvement of our proposed SessionRec over the best baseline. “R@K” and “N@K” stand for Recall@K and NDCG@K, respectively. The performance of SessionRec is reported with the backbone of HSTU.}
\label{tab:model_performance}
\resizebox{\textwidth}{!}{%
\begin{tabular}{@{}clcccccc|cccccccc@{}} 
\toprule 
\multirow{2.5}{*}{\textbf{Dataset}}  &\multirow{2.5}{*}{\textbf{Model}}
& \multicolumn{6}{c}{\textbf{Leave-One-Session-Out}} & \multicolumn{6}{|c}{\textbf{Leave-One-Item-Out}} \\
\cmidrule(lr){3-8} \cmidrule(lr){9-14}
&  & \textbf{N@10} & \textbf{N@100} & \textbf{N@500} & \textbf{R@10} & \textbf{R@100} & \textbf{R@500} & \textbf{N@10} & \textbf{N@100} & \textbf{N@500} & \textbf{R@10} & \textbf{R@100} & \textbf{R@500} \\
\midrule
\multirow{8}{*}{KuaiSar} 
     & GRU4Rec         & 0.0137 & 0.0209 & 0.0269 & 0.0186 & 0.0484 & 0.0903 & 0.0083 & 0.0141 & 0.0185 & 0.0140 & 0.0435 & 0.0780 \\
     & SASRec          & 0.0748 & 0.0814 & 0.0853 & 0.0818 & 0.1120 & 0.1447 & 0.0593 & 0.0655 & 0.0698 & 0.0824 & 0.1133 & 0.1473 \\
     & Bert4Rec        & 0.0413 & 0.0542 & 0.0615 & 0.0568 & 0.1119 & 0.1640 & 0.0325 & 0.0426 & 0.0487 & 0.0527 & 0.1029 & 0.1503 \\
     & SASRec+         & 0.0782 & 0.0862 & 0.0914 & 0.0893 & 0.1270 & 0.1662 & 0.0723 & 0.0790 & 0.0832 & 0.0939 & 0.1271 & 0.1605 \\
     & HSTU            & 0.0837 & 0.0956 & 0.1016 & 0.1004 & 0.1556 & 0.2009 & 0.0772 & 0.0859 & 0.0905 & 0.1019 & 0.1438 & 0.1800 \\
     & HSTU+           & \underline{0.0934} & \underline{0.1078} & \underline{0.1129} & \underline{0.1136} & \underline{0.1809} & \underline{0.2212} & \underline{0.0869} & \underline{0.0988} & \underline{0.1037} & \underline{0.1179} & \underline{0.1755} & \underline{0.2131} \\
     \addlinespace[1pt] 
     \cmidrule(lr){2-8} \cmidrule(lr){9-14}
     \addlinespace[1pt] 
     & SessionRec-HSTU & \textbf{0.0950} & \textbf{0.1180} & \textbf{0.1287} & \textbf{0.1268} & \textbf{0.2413} & \textbf{0.3318} & \textbf{0.0871} & \textbf{0.1076} & \textbf{0.1175} & \textbf{0.1323} & \textbf{0.2342} & \textbf{0.3107} \\
     & Impv.           & +1.71\% & +9.46\% & +13.99\% & +11.62\% & +33.39\% & +49.98\% & +0.23\% & +8.91\% & +13.30\% & +12.21\% & +33.45\% & +45.80\% \\
\midrule
\multirow{8}{*}{RecFlow} 
     & GRU4Rec         & 0.0039 & 0.0098 & 0.0190 & 0.0045 & 0.0214 & 0.0619 & 0.0027 & 0.0058 & 0.0108 & 0.0055 & 0.0222 & 0.0621 \\
     & SASRec          & 0.0021 & 0.0069 & 0.0143 & 0.0027 & 0.0163 & 0.0490 & 0.0005 & 0.0015 & 0.0033 & 0.0009 & 0.0065 & 0.0209 \\
     & Bert4Rec        & 0.0032 & 0.0095 & 0.0184 & 0.0034 & 0.0213 & 0.0614 & 0.0022 & 0.0057 & 0.0105 & 0.0038 & 0.0219 & 0.0602 \\
     & SASRec+         & 0.0042 & 0.0110 & 0.0202 & 0.0047 & 0.0241 & 0.0663 & 0.0030 & 0.0071 & 0.0126 & 0.0063 & 0.0282 & 0.0718 \\
     & HSTU            & 0.0054 & 0.0152 & 0.0271 & 0.0061 & 0.0349 & 0.0918 & 0.0033 & 0.0079 & 0.0143 & 0.0067 & 0.0304 & 0.0813 \\
     & HSTU+           & \underline{0.0061} & \underline{0.0158} & \underline{0.0283} & \underline{0.0074} & \underline{0.0353} & \underline{0.0949} & \underline{0.0038} & \underline{0.0085} & \underline{0.0151} & \underline{0.0075} & \underline{0.0323} & \underline{0.0850} \\
     \addlinespace[1pt] 
     \cmidrule(lr){2-8} \cmidrule(lr){9-14}
     \addlinespace[1pt] 
     & SessionRec-HSTU & \textbf{0.0082} & \textbf{0.0215} & \textbf{0.0372} & \textbf{0.0096} & \textbf{0.0495} & \textbf{0.1281} & \textbf{0.0051} & \textbf{0.0123} & \textbf{0.0224} & \textbf{0.0098} & \textbf{0.0479} & \textbf{0.1274} \\
     & Impv.           & +34.43\% & +36.08\% & +31.45\% & +29.73\% & +40.23\% & +34.98\% & +34.21\% & +44.71\% & +48.34\% & +30.67\% & +48.29\% & +49.88\% \\
\bottomrule
\end{tabular} 
}
\end{table*}

\subsubsection{Baselines}

To verify the effectiveness of our method, we compare it with the representative sequential recommendation baselines: GRU4Rec, SASRec, SASRec+, Bert4Rec, and the most recent strong generative recommendation model HSTU.
\begin{itemize}
\item \textbf{GRU4Rec} \cite{hidasi2018gru4rec}: An RNN-based model that uses GRU to model user behavior sequences. 
\item \textbf{SASRec} \cite{kang2018self}: The first and most classic transformer-based sequential recommendation model with unidirectional causal self-attention.
\item \textbf{SASRec+} \cite{klenitskiy2023turning}: An improved version of SAS4Rec,  which trained with negative sampling and cross-entropy loss.
\item \textbf{BERT4Rec} \cite{sun2019bert4rec}: A representative sequential recommendation model, which employs the deep bidirectional self-attention to model user behavior sequences. 
\item \textbf{HSTU} \cite{zhai24hstu}: A recent novel sequential transduction generative architecture designed for high cardinality, non-stationary recommendation data, which achieves large improvement over previous work and is a strong baseline model.
\item  \textbf{HSTU+} \cite{zhai24hstu}:  HSTU+ extends HSTU by including negative interactions in the input sequence, while HSTU in its open source code is limited to positive interactions.

\end{itemize}

\subsubsection{Evaluation Setting \& Metrics}
Previous works usually adopt an leave-one-item-out strategy for evaluation (i.e., next item prediction), which leaves the last item of the interaction sequence as test data \cite{kang2018self, sun2019bert4rec, tang2018personalized, klenitskiy2023turning}. We argue that this evaluation method is not applicable to real-world recommendation system scenarios, as users may interact with multiple items in the results of a single recommendation session/request, and items within the same session cannot serve as prior information for each other when evaluated. Therefore, we additionally adopt a leave-one-session-out strategy (i.e., next session prediction), using all items that the user positively interacted with in the last session as test data, which also matches the next session paradigm. For a fair comparison with previous work, we present the performance of different methods using both strategies in the experimental section. While we also suggest employing leave-one-session-out evaluation strategy in recommendation scenarios where multiple items are returned in a single session.

For evaluation metrics, we utilize two most widely adopted metrics: NDCG@\( K \)(N@\( K \)) and Recall@\( K \) (R@\( K \)) with \( K \)=10, 100, 500. And we report full unsampled metrics over the entire candidate items for our experiments, consistent with recent works \cite{zhai24hstu,zhai2023revisiting, klenitskiy2023turning}. For full unsampled metrics and datasets with a large number of items, we adopt a larger \( K \) \cite{klenitskiy2023turning,liu2024recflow}. For all these metrics, the higher the value, the better the performance.

\subsubsection{Implementation Details}
For KuaiSAR, we set the maximum length of positive interactions extracted from a sequence to 200. For RecFlow, the length is set to 400. We utilize a batch size of 64. All models are trained 200 epochs, with a learning rate of 0.001 and a dropout rate of 0.2. Throughout the training process, with the exception of SASRec, all other methods incorporate 128 negative samples per item. To ensure a fair comparison, the number of network layers for all models is set to 4, striking a balance between model complexity and computational feasibility. For Transformer-based models, attention heads is 2. We implement all models based on the open source code repository\footnote{\url{https://github.com/antklen/sasrec-bert4rec-recsys23}} in \cite{klenitskiy2023turning}. All baseline models, except for HSTU, have been implemented in this code repository. For HSTU, we migrate the core hstu block from the original paper's implement into the aforementioned code repository, ensuring that all models remain consistent in all aspects except for the modeling module. We will make our code of data processing, models, and configurations publicly available to promote reproducibility.

\subsection{Overall Performance Comparison}
In Table \ref{tab:model_performance}, we compare the performance of the proposed session-level models and other item-level models adopting both the leave-one-item-out and the leave-one-session-out strategy. SessionRec-HSTU significantly outperforms the baselines under these two configurations. By performing computations at the session level, SessionRec-HSTU predicts diverse interests across a broader space, which enables simultaneous enhancement of prediction accuracy at both session and item levels. HSTU-based models demonstrate the best performance, followed by Transformer-based models, with GRU-based models performing the worst. Increasing the complexity of the model can enhance the accuracy of item-level models to some extent. Additionally, SASRec+ surpassing SASRec indicates that training with a multiple negative sampling strategy effectively improves model accuracy, aligning with the findings in \cite{klenitskiy2023turning}.

\begin{figure}[t]
  \centering
  \includegraphics[width=0.9\linewidth]{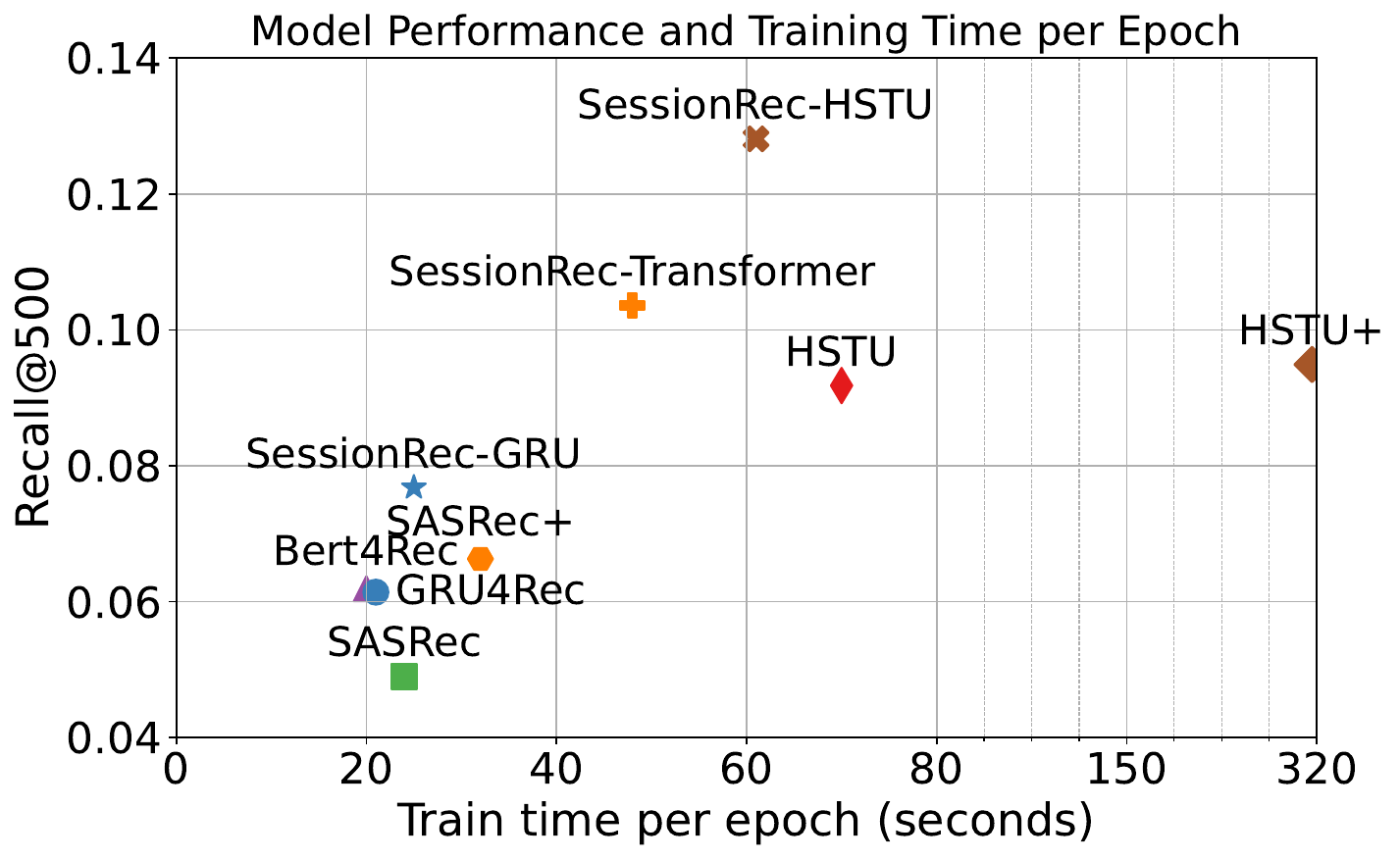}
  \vspace{-0.3cm}
  \caption{Performance and Traning Time per Epoch}
  \label{tab:performance_cost}
  \vspace{-0.3cm}
\end{figure}

Furthermore, the comparison between HSTU, relying exclusively on positive interactions, and HSTU+, which integrates extensive negative interaction behavior into the input sequence, demonstrates models benefit from incorporating full interactions. As depicted in Figure \ref{tab:performance_cost}, training on the same number of positive and negative interactions, SessionRec-HSTU significantly reduces training time compared to HSTU+ via session-level aggregation, which significantly lowers computational time and improves overall efficiency.
In conclusion, the results reveal that session-level models:
a) exhibit superior ability to capture multiple interests within a single session, enhancing both next-session and next-item prediction accuracy;
b) benefit from incorporating negative samples during training, boosting model robustness and generalization; and c) demonstrate notable computational efficiency improvements.

\subsection{Flexible Model-Agnostic Paradigm}
\begin{table}[t]
\caption{Model performance with improvement percentage.}
\vspace{-0.1cm}
\label{tab:model_impovement}
\resizebox{0.48\textwidth}{!}{
\centering 
\begin{tabular}{@{}lllll@{}} 
\toprule
Dataset & Model & N@500 & R@500 \\
\midrule
\multirow{6}{*}{\centering KuaiSar} & GRU & 0.0269 & 0.0903 \\
& SessionRec-GRU & 0.0394 (+46.47\%) & 0.1319 (+46.07\%) \\
\addlinespace[1pt] 
 \cline{2-4}
 \addlinespace[1pt] 
& SASRec+ & 0.0914 & 0.1662 \\
& SessionRec-Transformer & 0.1279 (+39.93\%) & 0.3279 (+97.29\%) \\
\addlinespace[1pt] 
 \cline{2-4}
 \addlinespace[1pt] 
& HSTU+ & 0.1129 & 0.2212 \\
& SessionRec-HSTU & 0.1287 (+13.99\%) & 0.3318 (+50.00\%) \\
\midrule
\multirow{6}{*}{\centering RecFlow} & GRU & 0.0190 & 0.0619 \\
& SessionRec-GRU & 0.0231 (+21.58\%) & 0.0768 (+24.07\%) \\
\addlinespace[1pt] 
 \cline{2-4}
 \addlinespace[1pt] 
& SASRec+ & 0.0202 & 0.0663 \\
& SessionRec-Transformer & 0.0304 (+50.50\%) & 0.1036 (+56.26\%) \\
\addlinespace[1pt] 
 \cline{2-4}
 \addlinespace[1pt] 
& HSTU+ & 0.0283 & 0.0949 \\
& SessionRec-HSTU & 0.0372 (+31.45\%) & 0.1281 (+34.98\%) \\
\bottomrule
\end{tabular}%
}
\vspace{-0.6cm}
\end{table}

Since our proposed method is model-agnostic with the characteristics of plug and play, we verify the performance of next session paradigm based on different backbone of GRU, Transformer and HSTU. Table \ref{tab:model_impovement} reports the results of NDCG@500 and Recall@500. The experimental results demonstrate that integrating the next session paradigm into different backbones can lead to substantial performance improvements over the original backbones.
On average, the performance enhancement across experimental datasets exceeds \textbf{39\%}. 
SessionRec-GRU and SessionRec-Transformer exhibit substantial gains in both NDCG@500 and Recall@500 metrics, highlighting the paradigm's ability to boost model accuracy. Notably, HSTU-based models, when augmented with session-level aggregation, consistently outperform their GRU and Transformer counterparts, emphasizing the paradigm's effectiveness in leveraging model complexity for superior results. This adaptability underscores the paradigm's potential to accommodate diverse application scenarios, making it a versatile tool for enhancing recommendation systems.

\subsection{Session Encoder Study}
\begin{table}[t]
\centering
\caption{SessionRec-HSTU aggregation comparison.}
\vspace{-0.1cm}
\label{tab:session_rec_hstu_aggregation}
\resizebox{0.48\textwidth}{!}{%
\begin{tabular}{@{}llcccccc@{}}
\toprule
Dataset & Aggregation & N@10 & N@100 & N@500 & R@10 & R@100 & R@500 \\
\midrule
\multirow{8}{*}{KuaiSar} 
    & mean         & \textbf{0.0950} & \textbf{0.1180} & \textbf{0.1287} & \textbf{0.1268} & \textbf{0.2413} & \textbf{0.3318} \\
    & max          & 0.0922 & 0.1149 & 0.1252 & 0.1235 & 0.2333 & 0.3217 \\
    & max\_relu    & 0.0852 & 0.1084 & 0.1187 & 0.1138 & 0.2267 & 0.3137 \\
    & gru(l=1)       & 0.0519 & 0.0683 & 0.0771 & 0.0714 & 0.1484 & 0.2195 \\
    & gru(l=2)       & 0.0872 & 0.0936 & 0.1040 & 0.1017 & 0.2090 & 0.2956 \\
    & trans(h=1,l=1) & 0.0797 & 0.1010 & 0.1115 & 0.1077 & 0.2142 & 0.3018 \\
    & trans(h=1,l=2) & 0.0708 & 0.0932 & 0.1034 & 0.0991 & 0.2076 & 0.2947 \\
    & trans(h=2,l=1) & 0.0561 & 0.0743 & 0.0832 & 0.0740 & 0.1577 & 0.2314 \\
    & trans(h=4,l=2) & 0.0537 & 0.0717 & 0.0807 & 0.0729 & 0.1574 & 0.2293 \\
\midrule
\multirow{8}{*}{RecFlow} 
    & mean         & \textbf{0.0082} & \textbf{0.0215} & \textbf{0.0372} & \textbf{0.0096} & \textbf{0.0495} & \textbf{0.1281} \\
    & max          & 0.0067 & 0.0178 & 0.0322 & 0.0077 & 0.0406 & 0.1111 \\
    & max\_relu    & 0.0062 & 0.0166 & 0.0305 & 0.0072 & 0.0390 & 0.1069 \\
    & gru(h=1)       & 0.0051 & 0.0147 & 0.0272 & 0.0058 & 0.0343 & 0.0952 \\
    & gru(h=2)       & 0.0045 & 0.0129 & 0.0253 & 0.0050 & 0.0305 & 0.0890 \\
    & trans(h=1,l=1) & 0.0054 & 0.0148 & 0.0283 & 0.0063 & 0.0342 & 0.0989 \\
    & trans(h=1,l=2) & 0.0053 & 0.0146 & 0.0278 & 0.0065 & 0.0343 & 0.0970 \\
    & trans(h=2,l=1) & 0.0045 & 0.0124 & 0.0245 & 0.0051 & 0.0287 & 0.0847 \\
    & trans(h=4,l=2) & 0.0051 & 0.0137 & 0.0258 & 0.0058 & 0.0319 & 0.0896 \\
\bottomrule
\end{tabular}%
}
\vspace{-0.3cm}
\end{table}

For the session encoder, we employ mean pooling(mean), max pooling(max), max pooling following a ReLU layer(max\_relu), GRU(gru), and Transformer(trans) to aggregate embeddings within the same session as shown in Table \ref{tab:session_rec_hstu_aggregation}, where h is the number of headers, and l is the number of layers. Specifically, among the tested methods, mean pooling yield the best performance, followed by max pooling and max pooling augmented with a ReLU layer. And sequential-based models exhibit comparatively weaker performance.

Upon further analysis, we observe that as computational complexity increases, performance generally tends to diminish. This suggests that during the aggregation phase, it may be beneficial to preserve the original representation of the items rather than introduce additional transformations. 
These findings imply that maintaining simplicity in representation helps capture essential session patterns more accurately, supporting the notion that straightforward aggregation techniques often suffice for effective session-based modeling.
\subsection{Rank Loss Study}
\begin{figure}[t]
  \centering
  \includegraphics[width=1.05\linewidth]{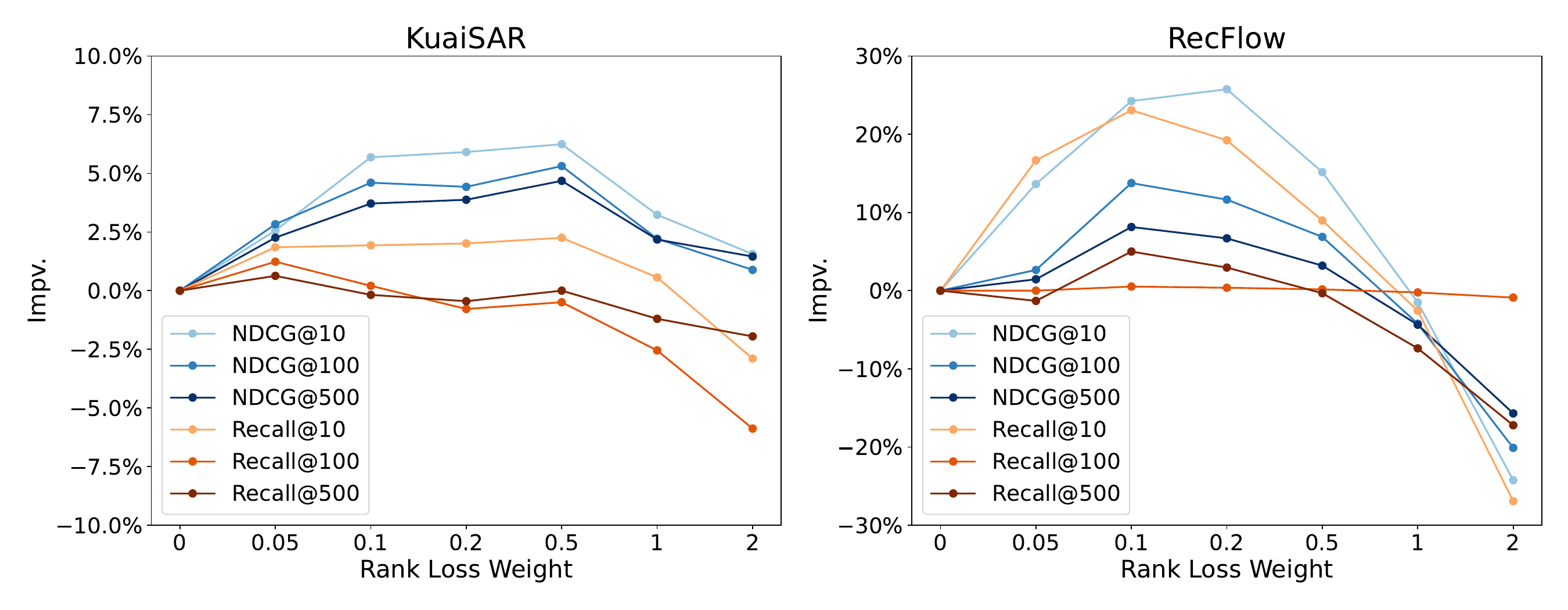}
    \vspace{-0.6cm}
  \caption{Impact of ranking loss weight on model performance improvement.}
  \label{tab:performance_rankloss}
  \vspace{-0.4cm}
\end{figure}
As illustrated in Figure \ref{tab:performance_rankloss}, during model training, we optimize and balance retrieval capabilities and ranking quality by combining rank loss and retrieval loss. Rank loss focuses on improving the intra-session ranking accuracy of items, while retrieval loss ensures the model effectively identifies positive samples. 
% This experiment aims to find the optimal rank loss weight to balance these two objectives.
We test various settings for rank loss weight $\alpha \in$ [0, 0.05, 0.1, 0.2, 0.5, 1, 2] and evaluate model performance on two datasets.
\begin{itemize} 
\item \textit{$\alpha$ = 0}: Utilizing only retrieval loss yields strong retrieval capability but suboptimal ranking performance. 
\item \textit{$\alpha$ = 0.2}: Achieves significant ranking accuracy improvements up to 6\% and 20\% on the two datasets respectively, while maintaining good retrieval coverage. 
\item \textit{$\alpha$ > 0.2}: Further increasing the weight continues to enhance ranking accuracy but adversely affects retrieval coverage, leading to an overall decline in performance. 
\end{itemize}

Specially, the pronounced fluctuations in Recall@10 compared to NDCG can be attributed to the rank accuracy enhancement, which drives the model to prioritize recommending items that are of top potential interest to the users. This makes Recall@10 a more precise measure of top-ranking results. In contrast, other recall metrics do not exhibit similar fluctuations with NDCG. When the rank loss weight is below 1, the recall performance remains relatively stable. However, when the weight exceeds 1, the overall performance in retrieval tasks deteriorates. 
These results underscore the importance of balancing rank loss and retrieval loss for optimal model performance. A moderate rank loss weight, such as 0.2 or 0.5, appears to offer the best trade-off, ensuring high ranking accuracy without compromising retrieval efficacy.

\section{Power-law Scaling Law of SessionRec}

\begin{figure}[t]
  \centering
  \includegraphics[width=0.9\linewidth]{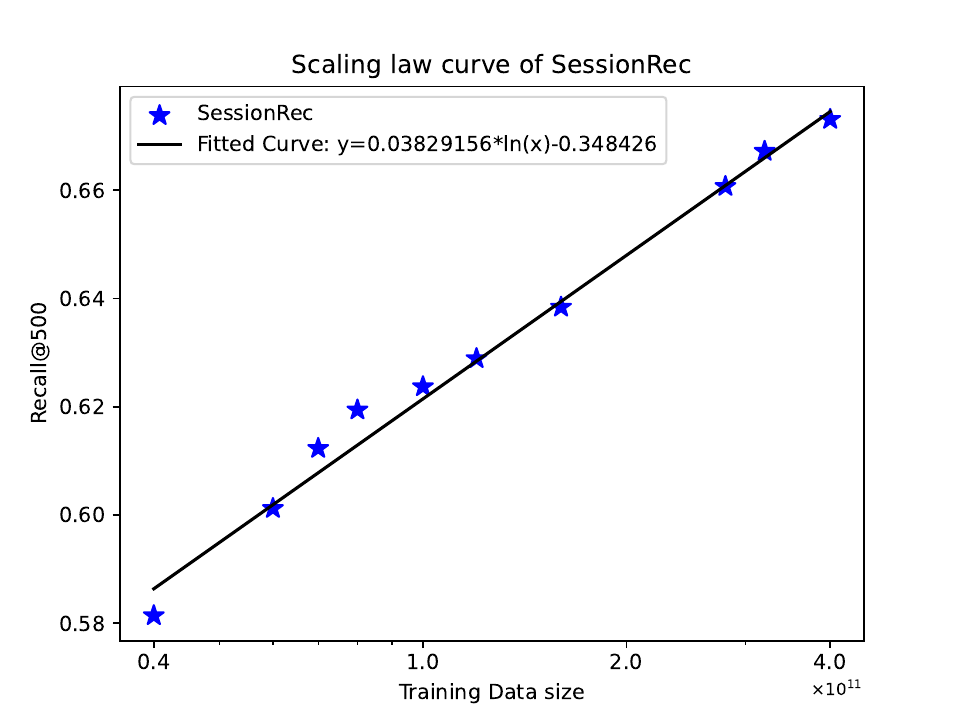}
    \vspace{-0.3cm}
  \caption{Scaling laws of SessionRec with data volume.}
    % \vspace{-0.5cm}
  \label{fig:scaling-law}

\end{figure}

We examined the scalability of SessionRec with HSTU backbone using a large-scale industrial dataset from the Meituan App, which contains comprehensive long-term user behavior sequences for approximately 400 million users. The results are illustrated in Figure \ref{fig:scaling-law}, where the horizontal axis denotes the training data size, quantified by the number of items in the training sequence. The vertical axis indicates the performance metric Recall@500, serving as the main offline evaluation indicator for retrieval. To prevent label leakage, user behavior sequences are ordered based on timestamps, employing the time span \([1, T]\) as the training window and the subsequent period \(T+1\) as the evaluation window.

It is evident from Figure \ref{fig:scaling-law} that with the exponential increase in training data volume and computing power, the performance of the model shows a trend of linear increase, indicating that the next session prediction paradigm still maintains great scalability, similar to those observed in LLMs.

\begin{figure}[t]
  % \vspace{-0.5cm}
  \centering
  \includegraphics[width=1.05\linewidth]{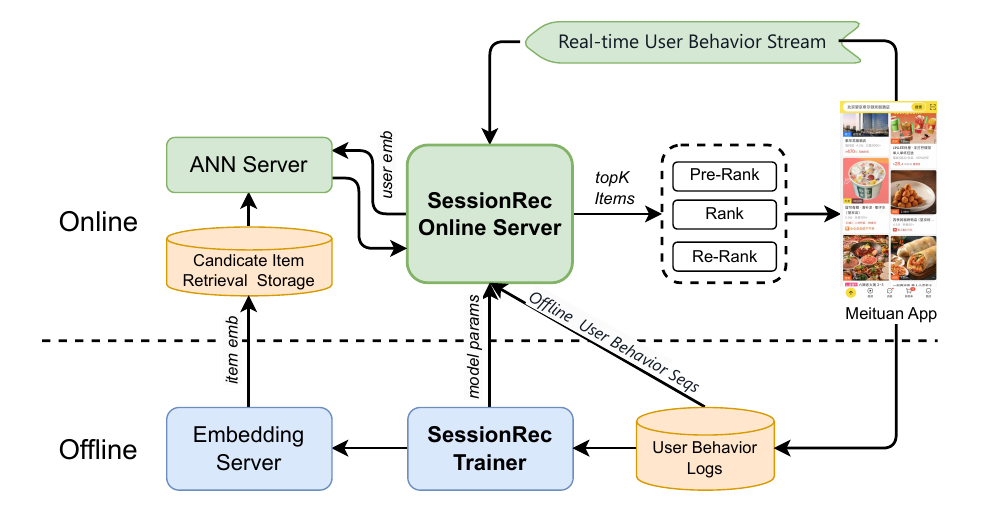}
    
  \caption{Deployment architecture of SessionRec in online recommendation system of Meituan App.}
  \vspace{-0.5cm}
  \label{fig:online_deploy}

\end{figure}

\section{Online Deployment and A/B Test}

Apart from offline experiments, SessionRec is also targeted at and successfully applied in real-world industrial practices. We adopt SessionRec with the backbone of HSTU in the online homepage recommendation system of Meituan App, which is one of the top-tier Apps in China and has more than 100 million active users per day. SessionRec is deployed in the retrieval stage, replacing the online SASRec model. As shown in Figure \ref{fig:online_deploy}, the input sequence of user behavior provided to the model consists of two parts: the offline sequence, derived from the system logs with the Spark tool, encapsulates user behaviors prior to the current day (\(T-1\) behaviors); and the real-time behavior stream that is captured using a near-line Flink computing engine. These two parts are combined by the online feature processing engine upon a user's request, forming the input sequence for the model. The maximum length of the input sequence is set to \(8K\). The model server takes the input sequence to generate user embedding, which represents the  user interest representation. Due to the large candidate pool and relatively stable features, \(T-1\)  offline computation is used to store item embeddings into the index database. The final topK retrieval results are obtained via efficient ANN algorithms (e.g. Faiss \cite{johnson2019billion}) and then transmitted to the recommendation system's downstream components.

We conducted \(A/B\) test for seven consecutive days, revealing that the deployment of SessionRec achieves significant improvements in both the quantity and efficiency of payments, with a 0.603\% increase in Page Views for Payment (Pay PV) and a 0. 564\% increase in  Page View Click-Through Conversion Rate (PVCTCVR). For Meituan, a life services platform, these metrics are the most important indicators for online observation. This demonstrates the practical value of SessionRec in industrial scenarios, as it effectively extracts user interests using historical behavior sequences, thereby boosting business metrics. We are continuously optimizing the real-time capabilities of the online server and will provide updates on the performance of the online deployment shortly.

\section{Conclusion and Future Work}

In this paper, we propose a novel generation paradigm called \textbf{SessionRec} that redefines the generative sequential recommendation as "next-session prediction", diverging from the conventional "next-item prediction". SessionRec more closely aligns with how users engage with real-world recommendation systems and stands out as a simple-yet-effective, plug-and-play methodology that makes it highly adaptable to any sequence modeling backbone. Furthermore, we propose a ranking loss that can significantly enhance the ranking performance of generative sequence recommendation models while simultaneously maintaining the retrieval performance. Extensive experiments conducted on public datasets and online A/B tests demonstrate the effectiveness of the proposed approach. For future work, we will explore whether a single generative sequence recommendation models, enhanced for improved ranking performance, can replace the multi-model cascaded paradigm of retrieval, pre-ranking, and ranking in traditional recommendation systems, and we believe this is a direction with considerable potential.

% \begin{acks}
% \end{acks}

\bibliographystyle{ACM-Reference-Format}
\bibliography{reference}


\begin{thebibliography}{43}

%%% the Web address.

\ifx \showCODEN    \undefined \def \showCODEN     #1{\unskip}     \fi
\ifx \showISBNx    \undefined \def \showISBNx     #1{\unskip}     \fi
\ifx \showISBNxiii \undefined \def \showISBNxiii  #1{\unskip}     \fi
\ifx \showISSN     \undefined \def \showISSN      #1{\unskip}     \fi
\ifx \showLCCN     \undefined \def \showLCCN      #1{\unskip}     \fi
\ifx \shownote     \undefined \def \shownote      #1{#1}          \fi
\ifx \showarticletitle \undefined \def \showarticletitle #1{#1}   \fi
\ifx \showURL      \undefined \def \showURL       {\relax}        \fi
% invisible to TeX
\providecommand\bibfield[2]{#2}
\providecommand\bibinfo[2]{#2}
\providecommand\natexlab[1]{#1}
\providecommand\showeprint[2][]{arXiv:#2}

\bibitem[Achiam et~al\mbox{.}(2023)]%
        {achiam2023gpt4}
\bibfield{author}{\bibinfo{person}{Josh Achiam}, \bibinfo{person}{Steven Adler}, \bibinfo{person}{Sandhini Agarwal}, \bibinfo{person}{Lama Ahmad}, \bibinfo{person}{Ilge Akkaya}, \bibinfo{person}{Florencia~Leoni Aleman}, \bibinfo{person}{Diogo Almeida}, \bibinfo{person}{Janko Altenschmidt}, \bibinfo{person}{Sam Altman}, \bibinfo{person}{Shyamal Anadkat}, {et~al\mbox{.}}} \bibinfo{year}{2023}\natexlab{}.
\newblock \showarticletitle{Gpt-4 technical report}.
\newblock \bibinfo{journal}{\emph{arXiv preprint arXiv:2303.08774}} (\bibinfo{year}{2023}).
\newblock


\bibitem[Brown et~al\mbox{.}(2020)]%
        {brown2020gpt3}
\bibfield{author}{\bibinfo{person}{Tom Brown}, \bibinfo{person}{Benjamin Mann}, \bibinfo{person}{Nick Ryder}, \bibinfo{person}{Melanie Subbiah}, \bibinfo{person}{Jared~D Kaplan}, \bibinfo{person}{Prafulla Dhariwal}, \bibinfo{person}{Arvind Neelakantan}, \bibinfo{person}{Pranav Shyam}, \bibinfo{person}{Girish Sastry}, \bibinfo{person}{Amanda Askell}, {et~al\mbox{.}}} \bibinfo{year}{2020}\natexlab{}.
\newblock \showarticletitle{Language models are few-shot learners}.
\newblock \bibinfo{journal}{\emph{Advances in neural information processing systems}}  \bibinfo{volume}{33} (\bibinfo{year}{2020}), \bibinfo{pages}{1877--1901}.
\newblock


\bibitem[Cao et~al\mbox{.}(2022)]%
        {cao2022sampling}
\bibfield{author}{\bibinfo{person}{Yue Cao}, \bibinfo{person}{Xiaojiang Zhou}, \bibinfo{person}{Jiaqi Feng}, \bibinfo{person}{Peihao Huang}, \bibinfo{person}{Yao Xiao}, \bibinfo{person}{Dayao Chen}, {and} \bibinfo{person}{Sheng Chen}.} \bibinfo{year}{2022}\natexlab{}.
\newblock \showarticletitle{Sampling is all you need on modeling long-term user behaviors for CTR prediction}. In \bibinfo{booktitle}{\emph{Proceedings of the 31st ACM International Conference on Information \& Knowledge Management}}. \bibinfo{pages}{2974--2983}.
\newblock


\bibitem[Chang et~al\mbox{.}(2023)]%
        {chang2023twin}
\bibfield{author}{\bibinfo{person}{Jianxin Chang}, \bibinfo{person}{Chenbin Zhang}, \bibinfo{person}{Zhiyi Fu}, \bibinfo{person}{Xiaoxue Zang}, \bibinfo{person}{Lin Guan}, \bibinfo{person}{Jing Lu}, \bibinfo{person}{Yiqun Hui}, \bibinfo{person}{Dewei Leng}, \bibinfo{person}{Yanan Niu}, \bibinfo{person}{Yang Song}, {et~al\mbox{.}}} \bibinfo{year}{2023}\natexlab{}.
\newblock \showarticletitle{TWIN: TWo-stage interest network for lifelong user behavior modeling in CTR prediction at kuaishou}. In \bibinfo{booktitle}{\emph{Proceedings of the 29th ACM SIGKDD Conference on Knowledge Discovery and Data Mining}}. \bibinfo{pages}{3785--3794}.
\newblock


\bibitem[Chen et~al\mbox{.}(2024)]%
        {chen2024hllm}
\bibfield{author}{\bibinfo{person}{Junyi Chen}, \bibinfo{person}{Lu Chi}, \bibinfo{person}{Bingyue Peng}, {and} \bibinfo{person}{Zehuan Yuan}.} \bibinfo{year}{2024}\natexlab{}.
\newblock \showarticletitle{Hllm: Enhancing sequential recommendations via hierarchical large language models for item and user modeling}.
\newblock \bibinfo{journal}{\emph{arXiv preprint arXiv:2409.12740}} (\bibinfo{year}{2024}).
\newblock


\bibitem[Chen et~al\mbox{.}(2021)]%
        {chen2021end}
\bibfield{author}{\bibinfo{person}{Qiwei Chen}, \bibinfo{person}{Changhua Pei}, \bibinfo{person}{Shanshan Lv}, \bibinfo{person}{Chao Li}, \bibinfo{person}{Junfeng Ge}, {and} \bibinfo{person}{Wenwu Ou}.} \bibinfo{year}{2021}\natexlab{}.
\newblock \showarticletitle{End-to-end user behavior retrieval in click-through rateprediction model}.
\newblock \bibinfo{journal}{\emph{arXiv preprint arXiv:2108.04468}} (\bibinfo{year}{2021}).
\newblock


\bibitem[Cheng et~al\mbox{.}(2016)]%
        {cheng2016wide}
\bibfield{author}{\bibinfo{person}{Heng-Tze Cheng}, \bibinfo{person}{Levent Koc}, \bibinfo{person}{Jeremiah Harmsen}, \bibinfo{person}{Tal Shaked}, \bibinfo{person}{Tushar Chandra}, \bibinfo{person}{Hrishi Aradhye}, \bibinfo{person}{Glen Anderson}, \bibinfo{person}{Greg Corrado}, \bibinfo{person}{Wei Chai}, \bibinfo{person}{Mustafa Ispir}, {et~al\mbox{.}}} \bibinfo{year}{2016}\natexlab{}.
\newblock \showarticletitle{Wide \& deep learning for recommender systems}. In \bibinfo{booktitle}{\emph{Proceedings of the 1st workshop on deep learning for recommender systems}}. \bibinfo{pages}{7--10}.
\newblock


\bibitem[Chung et~al\mbox{.}(2014)]%
        {chung2014gru}
\bibfield{author}{\bibinfo{person}{Junyoung Chung}, \bibinfo{person}{Caglar Gulcehre}, \bibinfo{person}{KyungHyun Cho}, {and} \bibinfo{person}{Yoshua Bengio}.} \bibinfo{year}{2014}\natexlab{}.
\newblock \showarticletitle{Empirical evaluation of gated recurrent neural networks on sequence modeling}.
\newblock \bibinfo{journal}{\emph{arXiv preprint arXiv:1412.3555}} (\bibinfo{year}{2014}).
\newblock


\bibitem[Doddapaneni et~al\mbox{.}(2024)]%
        {doddapaneni2024user}
\bibfield{author}{\bibinfo{person}{Sumanth Doddapaneni}, \bibinfo{person}{Krishna Sayana}, \bibinfo{person}{Ambarish Jash}, \bibinfo{person}{Sukhdeep Sodhi}, {and} \bibinfo{person}{Dima Kuzmin}.} \bibinfo{year}{2024}\natexlab{}.
\newblock \showarticletitle{User Embedding Model for Personalized Language Prompting}.
\newblock \bibinfo{journal}{\emph{arXiv preprint arXiv:2401.04858}} (\bibinfo{year}{2024}).
\newblock


\bibitem[Dosovitskiy(2020)]%
        {dosovitskiy2020image}
\bibfield{author}{\bibinfo{person}{Alexey Dosovitskiy}.} \bibinfo{year}{2020}\natexlab{}.
\newblock \showarticletitle{An image is worth 16x16 words: Transformers for image recognition at scale}.
\newblock \bibinfo{journal}{\emph{arXiv preprint arXiv:2010.11929}} (\bibinfo{year}{2020}).
\newblock


\bibitem[Feng et~al\mbox{.}(2019)]%
        {feng2019dsin}
\bibfield{author}{\bibinfo{person}{Yufei Feng}, \bibinfo{person}{Fuyu Lv}, \bibinfo{person}{Weichen Shen}, \bibinfo{person}{Menghan Wang}, \bibinfo{person}{Fei Sun}, \bibinfo{person}{Yu Zhu}, {and} \bibinfo{person}{Keping Yang}.} \bibinfo{year}{2019}\natexlab{}.
\newblock \showarticletitle{Deep session interest network for click-through rate prediction}.
\newblock \bibinfo{journal}{\emph{arXiv preprint arXiv:1905.06482}} (\bibinfo{year}{2019}).
\newblock


\bibitem[Harper and Konstan(2015)]%
        {harper2015movielens}
\bibfield{author}{\bibinfo{person}{F~Maxwell Harper} {and} \bibinfo{person}{Joseph~A Konstan}.} \bibinfo{year}{2015}\natexlab{}.
\newblock \showarticletitle{The movielens datasets: History and context}.
\newblock \bibinfo{journal}{\emph{Acm transactions on interactive intelligent systems (tiis)}} \bibinfo{volume}{5}, \bibinfo{number}{4} (\bibinfo{year}{2015}), \bibinfo{pages}{1--19}.
\newblock


\bibitem[Hidasi and Karatzoglou(2018)]%
        {hidasi2018gru4rec}
\bibfield{author}{\bibinfo{person}{Bal{\'a}zs Hidasi} {and} \bibinfo{person}{Alexandros Karatzoglou}.} \bibinfo{year}{2018}\natexlab{}.
\newblock \showarticletitle{Recurrent neural networks with top-k gains for session-based recommendations}. In \bibinfo{booktitle}{\emph{Proceedings of the 27th ACM international conference on information and knowledge management}}. \bibinfo{pages}{843--852}.
\newblock


\bibitem[Johnson et~al\mbox{.}(2019)]%
        {johnson2019billion}
\bibfield{author}{\bibinfo{person}{Jeff Johnson}, \bibinfo{person}{Matthijs Douze}, {and} \bibinfo{person}{Herv{\'e} J{\'e}gou}.} \bibinfo{year}{2019}\natexlab{}.
\newblock \showarticletitle{Billion-scale similarity search with {GPUs}}.
\newblock \bibinfo{journal}{\emph{IEEE Transactions on Big Data}} \bibinfo{volume}{7}, \bibinfo{number}{3} (\bibinfo{year}{2019}), \bibinfo{pages}{535--547}.
\newblock


\bibitem[Kang and McAuley(2018)]%
        {kang2018self}
\bibfield{author}{\bibinfo{person}{Wang-Cheng Kang} {and} \bibinfo{person}{Julian McAuley}.} \bibinfo{year}{2018}\natexlab{}.
\newblock \showarticletitle{Self-attentive sequential recommendation}. In \bibinfo{booktitle}{\emph{2018 IEEE international conference on data mining (ICDM)}}. IEEE, \bibinfo{pages}{197--206}.
\newblock


\bibitem[Kaplan et~al\mbox{.}(2020)]%
        {kaplan2020scaling}
\bibfield{author}{\bibinfo{person}{Jared Kaplan}, \bibinfo{person}{Sam McCandlish}, \bibinfo{person}{Tom Henighan}, \bibinfo{person}{Tom~B Brown}, \bibinfo{person}{Benjamin Chess}, \bibinfo{person}{Rewon Child}, \bibinfo{person}{Scott Gray}, \bibinfo{person}{Alec Radford}, \bibinfo{person}{Jeffrey Wu}, {and} \bibinfo{person}{Dario Amodei}.} \bibinfo{year}{2020}\natexlab{}.
\newblock \showarticletitle{Scaling laws for neural language models}.
\newblock \bibinfo{journal}{\emph{arXiv preprint arXiv:2001.08361}} (\bibinfo{year}{2020}).
\newblock


\bibitem[Klenitskiy and Vasilev(2023)]%
        {klenitskiy2023turning}
\bibfield{author}{\bibinfo{person}{Anton Klenitskiy} {and} \bibinfo{person}{Alexey Vasilev}.} \bibinfo{year}{2023}\natexlab{}.
\newblock \showarticletitle{Turning Dross Into Gold Loss: is BERT4Rec really better than SASRec?}. In \bibinfo{booktitle}{\emph{Proceedings of the 17th ACM Conference on Recommender Systems}}. \bibinfo{pages}{1120--1125}.
\newblock


\bibitem[Lee et~al\mbox{.}(2022)]%
        {lee2022autoregressive}
\bibfield{author}{\bibinfo{person}{Doyup Lee}, \bibinfo{person}{Chiheon Kim}, \bibinfo{person}{Saehoon Kim}, \bibinfo{person}{Minsu Cho}, {and} \bibinfo{person}{Wook-Shin Han}.} \bibinfo{year}{2022}\natexlab{}.
\newblock \showarticletitle{Autoregressive image generation using residual quantization}. In \bibinfo{booktitle}{\emph{Proceedings of the IEEE/CVF Conference on Computer Vision and Pattern Recognition}}. \bibinfo{pages}{11523--11532}.
\newblock


\bibitem[Lian et~al\mbox{.}(2018)]%
        {lian2018xdeepfm}
\bibfield{author}{\bibinfo{person}{Jianxun Lian}, \bibinfo{person}{Xiaohuan Zhou}, \bibinfo{person}{Fuzheng Zhang}, \bibinfo{person}{Zhongxia Chen}, \bibinfo{person}{Xing Xie}, {and} \bibinfo{person}{Guangzhong Sun}.} \bibinfo{year}{2018}\natexlab{}.
\newblock \showarticletitle{xdeepfm: Combining explicit and implicit feature interactions for recommender systems}. In \bibinfo{booktitle}{\emph{Proceedings of the 24th ACM SIGKDD international conference on knowledge discovery \& data mining}}. \bibinfo{pages}{1754--1763}.
\newblock


\bibitem[Liu et~al\mbox{.}(1998)]%
        {liu1998integrating}
\bibfield{author}{\bibinfo{person}{Bing Liu}, \bibinfo{person}{Wynne Hsu}, {and} \bibinfo{person}{Yiming Ma}.} \bibinfo{year}{1998}\natexlab{}.
\newblock \showarticletitle{Integrating classification and association rule mining}. In \bibinfo{booktitle}{\emph{Proceedings of the fourth international conference on knowledge discovery and data mining}}. \bibinfo{pages}{80--86}.
\newblock


\bibitem[Liu et~al\mbox{.}(2024)]%
        {liu2024recflow}
\bibfield{author}{\bibinfo{person}{Qi Liu}, \bibinfo{person}{Kai Zheng}, \bibinfo{person}{Rui Huang}, \bibinfo{person}{Wuchao Li}, \bibinfo{person}{Kuo Cai}, \bibinfo{person}{Yuan Chai}, \bibinfo{person}{Yanan Niu}, \bibinfo{person}{Yiqun Hui}, \bibinfo{person}{Bing Han}, \bibinfo{person}{Na Mou}, {et~al\mbox{.}}} \bibinfo{year}{2024}\natexlab{}.
\newblock \showarticletitle{RecFlow: An Industrial Full Flow Recommendation Dataset}.
\newblock \bibinfo{journal}{\emph{arXiv preprint arXiv:2410.20868}} (\bibinfo{year}{2024}).
\newblock


\bibitem[Liu et~al\mbox{.}(2021)]%
        {liu2021swin}
\bibfield{author}{\bibinfo{person}{Ze Liu}, \bibinfo{person}{Yutong Lin}, \bibinfo{person}{Yue Cao}, \bibinfo{person}{Han Hu}, \bibinfo{person}{Yixuan Wei}, \bibinfo{person}{Zheng Zhang}, \bibinfo{person}{Stephen Lin}, {and} \bibinfo{person}{Baining Guo}.} \bibinfo{year}{2021}\natexlab{}.
\newblock \showarticletitle{Swin transformer: Hierarchical vision transformer using shifted windows}. In \bibinfo{booktitle}{\emph{Proceedings of the IEEE/CVF international conference on computer vision}}. \bibinfo{pages}{10012--10022}.
\newblock


\bibitem[Malkov and Yashunin(2018)]%
        {malkov2018ann}
\bibfield{author}{\bibinfo{person}{Yu~A Malkov} {and} \bibinfo{person}{Dmitry~A Yashunin}.} \bibinfo{year}{2018}\natexlab{}.
\newblock \showarticletitle{Efficient and robust approximate nearest neighbor search using hierarchical navigable small world graphs}.
\newblock \bibinfo{journal}{\emph{IEEE transactions on pattern analysis and machine intelligence}} \bibinfo{volume}{42}, \bibinfo{number}{4} (\bibinfo{year}{2018}), \bibinfo{pages}{824--836}.
\newblock


\bibitem[McAuley et~al\mbox{.}(2015)]%
        {mcauley2015amazon}
\bibfield{author}{\bibinfo{person}{Julian McAuley}, \bibinfo{person}{Christopher Targett}, \bibinfo{person}{Qinfeng Shi}, {and} \bibinfo{person}{Anton Van Den~Hengel}.} \bibinfo{year}{2015}\natexlab{}.
\newblock \showarticletitle{Image-based recommendations on styles and substitutes}. In \bibinfo{booktitle}{\emph{Proceedings of the 38th international ACM SIGIR conference on research and development in information retrieval}}. \bibinfo{pages}{43--52}.
\newblock


\bibitem[Ouyang et~al\mbox{.}(2022)]%
        {ouyang2022training}
\bibfield{author}{\bibinfo{person}{Long Ouyang}, \bibinfo{person}{Jeffrey Wu}, \bibinfo{person}{Xu Jiang}, \bibinfo{person}{Diogo Almeida}, \bibinfo{person}{Carroll Wainwright}, \bibinfo{person}{Pamela Mishkin}, \bibinfo{person}{Chong Zhang}, \bibinfo{person}{Sandhini Agarwal}, \bibinfo{person}{Katarina Slama}, \bibinfo{person}{Alex Ray}, {et~al\mbox{.}}} \bibinfo{year}{2022}\natexlab{}.
\newblock \showarticletitle{Training language models to follow instructions with human feedback}.
\newblock \bibinfo{journal}{\emph{Advances in neural information processing systems}}  \bibinfo{volume}{35} (\bibinfo{year}{2022}), \bibinfo{pages}{27730--27744}.
\newblock


\bibitem[Pi et~al\mbox{.}(2019)]%
        {pi2019mimn}
\bibfield{author}{\bibinfo{person}{Qi Pi}, \bibinfo{person}{Weijie Bian}, \bibinfo{person}{Guorui Zhou}, \bibinfo{person}{Xiaoqiang Zhu}, {and} \bibinfo{person}{Kun Gai}.} \bibinfo{year}{2019}\natexlab{}.
\newblock \showarticletitle{Practice on long sequential user behavior modeling for click-through rate prediction}. In \bibinfo{booktitle}{\emph{Proceedings of the 25th ACM SIGKDD International Conference on Knowledge Discovery \& Data Mining}}. \bibinfo{pages}{2671--2679}.
\newblock


\bibitem[Pi et~al\mbox{.}(2020)]%
        {pi2020search}
\bibfield{author}{\bibinfo{person}{Qi Pi}, \bibinfo{person}{Guorui Zhou}, \bibinfo{person}{Yujing Zhang}, \bibinfo{person}{Zhe Wang}, \bibinfo{person}{Lejian Ren}, \bibinfo{person}{Ying Fan}, \bibinfo{person}{Xiaoqiang Zhu}, {and} \bibinfo{person}{Kun Gai}.} \bibinfo{year}{2020}\natexlab{}.
\newblock \showarticletitle{Search-based user interest modeling with lifelong sequential behavior data for click-through rate prediction}. In \bibinfo{booktitle}{\emph{Proceedings of the 29th ACM International Conference on Information \& Knowledge Management}}. \bibinfo{pages}{2685--2692}.
\newblock


\bibitem[Radford(2018)]%
        {radford2018gpt1}
\bibfield{author}{\bibinfo{person}{Alec Radford}.} \bibinfo{year}{2018}\natexlab{}.
\newblock \showarticletitle{Improving language understanding by generative pre-training}.
\newblock  (\bibinfo{year}{2018}).
\newblock


\bibitem[Radford et~al\mbox{.}(2019)]%
        {radford2019gpt2}
\bibfield{author}{\bibinfo{person}{Alec Radford}, \bibinfo{person}{Jeffrey Wu}, \bibinfo{person}{Rewon Child}, \bibinfo{person}{David Luan}, \bibinfo{person}{Dario Amodei}, \bibinfo{person}{Ilya Sutskever}, {et~al\mbox{.}}} \bibinfo{year}{2019}\natexlab{}.
\newblock \showarticletitle{Language models are unsupervised multitask learners}.
\newblock \bibinfo{journal}{\emph{OpenAI blog}} \bibinfo{volume}{1}, \bibinfo{number}{8} (\bibinfo{year}{2019}), \bibinfo{pages}{9}.
\newblock


\bibitem[Rajput et~al\mbox{.}(2023)]%
        {rajput2023recommender}
\bibfield{author}{\bibinfo{person}{Shashank Rajput}, \bibinfo{person}{Nikhil Mehta}, \bibinfo{person}{Anima Singh}, \bibinfo{person}{Raghunandan Hulikal~Keshavan}, \bibinfo{person}{Trung Vu}, \bibinfo{person}{Lukasz Heldt}, \bibinfo{person}{Lichan Hong}, \bibinfo{person}{Yi Tay}, \bibinfo{person}{Vinh Tran}, \bibinfo{person}{Jonah Samost}, {et~al\mbox{.}}} \bibinfo{year}{2023}\natexlab{}.
\newblock \showarticletitle{Recommender systems with generative retrieval}.
\newblock \bibinfo{journal}{\emph{Advances in Neural Information Processing Systems}}  \bibinfo{volume}{36} (\bibinfo{year}{2023}), \bibinfo{pages}{10299--10315}.
\newblock


\bibitem[Song et~al\mbox{.}(2019)]%
        {song2019autoint}
\bibfield{author}{\bibinfo{person}{Weiping Song}, \bibinfo{person}{Chence Shi}, \bibinfo{person}{Zhiping Xiao}, \bibinfo{person}{Zhijian Duan}, \bibinfo{person}{Yewen Xu}, \bibinfo{person}{Ming Zhang}, {and} \bibinfo{person}{Jian Tang}.} \bibinfo{year}{2019}\natexlab{}.
\newblock \showarticletitle{Autoint: Automatic feature interaction learning via self-attentive neural networks}. In \bibinfo{booktitle}{\emph{Proceedings of the 28th ACM international conference on information and knowledge management}}. \bibinfo{pages}{1161--1170}.
\newblock


\bibitem[Sun et~al\mbox{.}(2019)]%
        {sun2019bert4rec}
\bibfield{author}{\bibinfo{person}{Fei Sun}, \bibinfo{person}{Jun Liu}, \bibinfo{person}{Jian Wu}, \bibinfo{person}{Changhua Pei}, \bibinfo{person}{Xiao Lin}, \bibinfo{person}{Wenwu Ou}, {and} \bibinfo{person}{Peng Jiang}.} \bibinfo{year}{2019}\natexlab{}.
\newblock \showarticletitle{BERT4Rec: Sequential recommendation with bidirectional encoder representations from transformer}. In \bibinfo{booktitle}{\emph{Proceedings of the 28th ACM international conference on information and knowledge management}}. \bibinfo{pages}{1441--1450}.
\newblock


\bibitem[Tang and Wang(2018)]%
        {tang2018personalized}
\bibfield{author}{\bibinfo{person}{Jiaxi Tang} {and} \bibinfo{person}{Ke Wang}.} \bibinfo{year}{2018}\natexlab{}.
\newblock \showarticletitle{Personalized top-n sequential recommendation via convolutional sequence embedding}. In \bibinfo{booktitle}{\emph{Proceedings of the eleventh ACM international conference on web search and data mining}}. \bibinfo{pages}{565--573}.
\newblock


\bibitem[Waswani et~al\mbox{.}(2017)]%
        {waswani2017transformer}
\bibfield{author}{\bibinfo{person}{A Waswani}, \bibinfo{person}{N Shazeer}, \bibinfo{person}{N Parmar}, \bibinfo{person}{J Uszkoreit}, \bibinfo{person}{L Jones}, \bibinfo{person}{A Gomez}, \bibinfo{person}{L Kaiser}, {and} \bibinfo{person}{I Polosukhin}.} \bibinfo{year}{2017}\natexlab{}.
\newblock \showarticletitle{Attention is all you need}. In \bibinfo{booktitle}{\emph{NIPS}}.
\newblock


\bibitem[Wei et~al\mbox{.}(2021)]%
        {wei2021finetuned}
\bibfield{author}{\bibinfo{person}{Jason Wei}, \bibinfo{person}{Maarten Bosma}, \bibinfo{person}{Vincent~Y Zhao}, \bibinfo{person}{Kelvin Guu}, \bibinfo{person}{Adams~Wei Yu}, \bibinfo{person}{Brian Lester}, \bibinfo{person}{Nan Du}, \bibinfo{person}{Andrew~M Dai}, {and} \bibinfo{person}{Quoc~V Le}.} \bibinfo{year}{2021}\natexlab{}.
\newblock \showarticletitle{Finetuned language models are zero-shot learners}.
\newblock \bibinfo{journal}{\emph{arXiv preprint arXiv:2109.01652}} (\bibinfo{year}{2021}).
\newblock


\bibitem[Wu et~al\mbox{.}(2017)]%
        {wu2017recurrent}
\bibfield{author}{\bibinfo{person}{Chao-Yuan Wu}, \bibinfo{person}{Amr Ahmed}, \bibinfo{person}{Alex Beutel}, \bibinfo{person}{Alexander~J Smola}, {and} \bibinfo{person}{How Jing}.} \bibinfo{year}{2017}\natexlab{}.
\newblock \showarticletitle{Recurrent recommender networks}. In \bibinfo{booktitle}{\emph{Proceedings of the tenth ACM international conference on web search and data mining}}. \bibinfo{pages}{495--503}.
\newblock


\bibitem[Yang et~al\mbox{.}(2024)]%
        {yang2024item}
\bibfield{author}{\bibinfo{person}{Li Yang}, \bibinfo{person}{Anushya Subbiah}, \bibinfo{person}{Hardik Patel}, \bibinfo{person}{Judith~Yue Li}, \bibinfo{person}{Yanwei Song}, \bibinfo{person}{Reza Mirghaderi}, {and} \bibinfo{person}{Vikram Aggarwal}.} \bibinfo{year}{2024}\natexlab{}.
\newblock \showarticletitle{Item-Language Model for Conversational Recommendation}.
\newblock \bibinfo{journal}{\emph{arXiv preprint arXiv:2406.02844}} (\bibinfo{year}{2024}).
\newblock


\bibitem[Yu et~al\mbox{.}(2019)]%
        {yu2019review}
\bibfield{author}{\bibinfo{person}{Yong Yu}, \bibinfo{person}{Xiaosheng Si}, \bibinfo{person}{Changhua Hu}, {and} \bibinfo{person}{Jianxun Zhang}.} \bibinfo{year}{2019}\natexlab{}.
\newblock \showarticletitle{A review of recurrent neural networks: LSTM cells and network architectures}.
\newblock \bibinfo{journal}{\emph{Neural computation}} \bibinfo{volume}{31}, \bibinfo{number}{7} (\bibinfo{year}{2019}), \bibinfo{pages}{1235--1270}.
\newblock


\bibitem[Zhai et~al\mbox{.}(2023)]%
        {zhai2023revisiting}
\bibfield{author}{\bibinfo{person}{Jiaqi Zhai}, \bibinfo{person}{Zhaojie Gong}, \bibinfo{person}{Yueming Wang}, \bibinfo{person}{Xiao Sun}, \bibinfo{person}{Zheng Yan}, \bibinfo{person}{Fu Li}, {and} \bibinfo{person}{Xing Liu}.} \bibinfo{year}{2023}\natexlab{}.
\newblock \showarticletitle{Revisiting Neural Retrieval on Accelerators}. In \bibinfo{booktitle}{\emph{Proceedings of the 29th ACM SIGKDD Conference on Knowledge Discovery and Data Mining}}. \bibinfo{pages}{5520--5531}.
\newblock


\bibitem[Zhai et~al\mbox{.}(2024)]%
        {zhai24hstu}
\bibfield{author}{\bibinfo{person}{Jiaqi Zhai}, \bibinfo{person}{Lucy Liao}, \bibinfo{person}{Xing Liu}, \bibinfo{person}{Yueming Wang}, \bibinfo{person}{Rui Li}, \bibinfo{person}{Xuan Cao}, \bibinfo{person}{Leon Gao}, \bibinfo{person}{Zhaojie Gong}, \bibinfo{person}{Fangda Gu}, \bibinfo{person}{Jiayuan He}, \bibinfo{person}{Yinghai Lu}, {and} \bibinfo{person}{Yu Shi}.} \bibinfo{year}{2024}\natexlab{}.
\newblock \showarticletitle{Actions Speak Louder than Words: Trillion-Parameter Sequential Transducers for Generative Recommendations}. In \bibinfo{booktitle}{\emph{Proceedings of the 41st International Conference on Machine Learning}} \emph{(\bibinfo{series}{Proceedings of Machine Learning Research}, Vol.~\bibinfo{volume}{235})}. \bibinfo{publisher}{PMLR}, \bibinfo{pages}{58484--58509}.
\newblock


\bibitem[Zhang et~al\mbox{.}(1996)]%
        {zhang1996birch}
\bibfield{author}{\bibinfo{person}{Tian Zhang}, \bibinfo{person}{Raghu Ramakrishnan}, {and} \bibinfo{person}{Miron Livny}.} \bibinfo{year}{1996}\natexlab{}.
\newblock \showarticletitle{BIRCH: an efficient data clustering method for very large databases}.
\newblock \bibinfo{journal}{\emph{ACM sigmod record}} \bibinfo{volume}{25}, \bibinfo{number}{2} (\bibinfo{year}{1996}), \bibinfo{pages}{103--114}.
\newblock


\bibitem[Zhou et~al\mbox{.}(2019)]%
        {zhou2019dien}
\bibfield{author}{\bibinfo{person}{Guorui Zhou}, \bibinfo{person}{Na Mou}, \bibinfo{person}{Ying Fan}, \bibinfo{person}{Qi Pi}, \bibinfo{person}{Weijie Bian}, \bibinfo{person}{Chang Zhou}, \bibinfo{person}{Xiaoqiang Zhu}, {and} \bibinfo{person}{Kun Gai}.} \bibinfo{year}{2019}\natexlab{}.
\newblock \showarticletitle{Deep interest evolution network for click-through rate prediction}. In \bibinfo{booktitle}{\emph{Proceedings of the AAAI conference on artificial intelligence}}, Vol.~\bibinfo{volume}{33}. \bibinfo{pages}{5941--5948}.
\newblock


\bibitem[Zhou et~al\mbox{.}(2018)]%
        {zhou2018din}
\bibfield{author}{\bibinfo{person}{Guorui Zhou}, \bibinfo{person}{Xiaoqiang Zhu}, \bibinfo{person}{Chenru Song}, \bibinfo{person}{Ying Fan}, \bibinfo{person}{Han Zhu}, \bibinfo{person}{Xiao Ma}, \bibinfo{person}{Yanghui Yan}, \bibinfo{person}{Junqi Jin}, \bibinfo{person}{Han Li}, {and} \bibinfo{person}{Kun Gai}.} \bibinfo{year}{2018}\natexlab{}.
\newblock \showarticletitle{Deep interest network for click-through rate prediction}. In \bibinfo{booktitle}{\emph{Proceedings of the 24th ACM SIGKDD international conference on knowledge discovery \& data mining}}. \bibinfo{pages}{1059--1068}.
\newblock


\end{thebibliography}

\appendix


\end{document}